\documentclass[a4paper,twoside,14pt]{article}

\usepackage[utf8]{inputenc}

\usepackage[english]{babel}
\usepackage{subeqnarray}
\usepackage{url}

\usepackage{graphicx,psfrag}
\usepackage{fixltx2e,amsmath}
\usepackage{xcolor}
\usepackage{amssymb}
\usepackage[final]{pdfpages}
\usepackage{bbold}
\usepackage{dsfont}
\usepackage{pgfplots}
\usepackage{subfig}
\usepackage{algorithm}
\usepackage{algorithmic,placeins}
\usepackage[font=small,labelfont=bf]{caption}

\usepackage[font={footnotesize}]{caption}

\usepackage[utf8]{inputenc}
\usepackage[T1]{fontenc}
\usepackage{xcolor,amsmath,amssymb}
\usepackage{geometry}
\usepackage{graphicx}
\usepackage[sort&compress,numbers]{natbib}
\usepackage{enumerate}
\usepackage{wasysym}
\usepackage[textsize=footnotesize]{todonotes}

\usepackage{hyperref}
\hypersetup{
    colorlinks=true,
    linkcolor=red,
    filecolor=magenta,      
    urlcolor=cyan,
}

\usepackage{pgfplots}

\definecolor{darkblue}{rgb}{0,0,0.6}
\definecolor{darkred}{rgb}{0.6,0,0}

\usepackage{hyperref}

\def\to{\rightarrow}

\newcommand{\beq}{\begin{equation}} \newcommand{\eeq}{\end{equation}}

\newcommand\be{\begin{equation}}
\newcommand\bea{\begin{eqnarray} \nonumber }
\newcommand\ee{\end{equation}}
\newcommand\eea{\end{eqnarray}}

\usepackage{authblk}

\usepackage[bitstream-charter]{mathdesign}

\usetikzlibrary{pgfplots.groupplots}

\usepackage{xcolor}

\graphicspath{{figures/}} 

\begin{document}

\title{The RFOT Theory of Glasses:\\ Recent Progress and Open Issues}
\author[1]{Giulio Biroli*\footnote{Email corresponding author: giulio.biroli@ens.fr}}
\author[2]{Jean-Philippe Bouchaud}
\affil[1]{Laboratoire de Physique de l’Ecole Normale Sup\'erieure, ENS, Universit\'e PSL, CNRS, Sorbonne Universit\'e, Universit\'e Paris Cit\'e, F-75005 Paris, France}
		\affil[2]{CFM, 23 rue de l'Universit\'e, F-75007 Paris, France,
		\& Acad\'emie des Sciences, Quai de Conti, F-75006 Paris, France}
		
\date{June 2022}
\maketitle

\abstract{The Random First Order Transition (RFOT) theory started with the pioneering 
work of Kirkpatrick, Thirumalai and Wolynes. It leverages the methods and advances of the theory of disordered systems. It fares remarkably well at reproducing the salient experimental facts of super-cooled liquids. 
Yet, direct and indisputable experimental validations are missing. 
In this short survey, we will review recent investigations that broadly support all static aspects of RFOT, but also those for which the standard dynamical extension of the theory appears to be struggling, in particular in relation with facilitation effects. We discuss possible solutions and open issues. } 

\section{Amorphous Order \& Random First Order Transitions}

As is well known, glasses are ``half-liquid, half-solid''. A glass has a structure factor characteristic of a liquid, yet glasses do not flow and respond elastically to shear deformations -- at least on time scales much shorter than the relaxation time $\tau$ of the system, which itself increases extraordinarily fast as temperature is decreased \cite{bbreview}. In the case of Ortho-Terphenyl, for example, the relaxation time increases by a factor $10^{10}$ as temperature drops by a mere $10\%$. 

From a general point of view, a non-zero static shear modulus is necessarily associated with a loss of ergodicity, and thus a transition into a state where the dynamics is no longer able to probe the entire phase-space \cite{yoshino}. The fundamental question that has riveted theoreticians for decades is whether the physics of glasses is indeed driven by an underlying phase transition into an ergodicity-broken state characterized by some amorphous long-range order, or whether the dramatic slowdown is purely of kinetic origin, with no particular thermodynamic signature. 

The concept of ``amorphous order'' sounds like an oxymoron, but accurately describes the physics of spin-glasses. Indeed, below some critical temperature $T_c$, each spin points in a random direction, but this direction remains fixed in time. Much as in glasses, instantaneous snapshots of the spin configurations seem featureless both above and below $T_c$. But whereas there is no long range transmission of information above $T_c$, the spin-glass phase is \textit{rigid}, in the sense that localised perturbations have a long range effect on the system -- much like the free-energy per particle of rigid bodies depends on the shape of its boundaries.

How much of the physics of spin-glasses, where a true thermodynamic transition takes place, is shared by super-cooled liquids and other glassy materials? From a theoretical point of view, the deep analogy between glasses and spin-glasses finds its roots in the landmark series of papers by Kirkpatrick, Thirumalai and Wolynes in the mid 80's \cite{KTW1,KTW2,KTW3}. Based on the solution of a family of mean-field models of spin-glasses, these authors proposed the ``Random First Order Transition'' (RFOT) theory \cite{KTW3,RFOT1,RFOT2}, which appears to capture all the known phenomenology of super-cooled liquids, in particular:
\begin{itemize}
    \item The existence of a cross-over temperature $T^\star$ below which metastable states appear that ``trap” the system for some large amount of time. This leads to a plateau in the relaxation function, associated with the appearance of local rigidity (also often called ``cage formation'') with a non-zero high frequency shear modulus $G_{\text{hf}}$.
    \item Such metastable states are exponentially numerous, with an associated extensive configurational entropy, $\Sigma(T)$. [The mere existence of such a large number of metastable states allows the system to decorrelate with time].
    \item The configurational entropy appears to vanish when the temperature is further lowered towards the Kauzmann temperature, which would correspond to a true thermodynamic (ideal glass) transition.  
    \item An Adam-Gibbs-like correlation between the logarithm of the relaxation time and the inverse of the configurational entropy, and between the fragility of the liquid and the jump of specific heat at the glass transition. 
\end{itemize} 

More precisely, the RFOT theory envisages the glass state as a mosaic of ``glassites'' (i.e. locally frozen clusters), resulting from a competition between extensive entropy of locally stable arrangements of molecules and the mismatch energy between two such configurations \cite{KTW3,Bou04}. The size $\ell$ of these glassites -- also called point-to-set \cite{Montanari-Semerjian} -- is then found to be inversely related to the configurational entropy and diverges as $T \downarrow T_K$  -- see Eq. \eqref{eq:ell_pts} below. Being of finite size, the life-time $\tau$ of these glassites is also finite, but grows exponentially with $\ell$ and, hence, diverges at $T_K$. These glassites are rigid, in the sense that boundary conditions are able to lock all inside molecules around a fixed position \cite{Bou04,Cavagna}. Hence, glassites respond elastically to an external shear for times less than $\tau$, with a shear modulus $G_{\text{hf}}$, but start flowing for times larger than $\tau$ when the local (amorphous) order finally unravels. 

RFOT theory is certainly one of the theory that fares best in terms of reproducing the phenomenology and salient experimental facts of super-cooled liquids, as extensively reviewed in \cite{wolynes-review,RFOT2}. Still, due to the relatively small length-scales at play, it is difficult for experiments to (yet?) provide direct and very stringent tests of the theory. 

On this front, numerical simulations have played a very important role in the last twenty years. Some aspects of the RFOT theory have received full confirmation from atomistic simulations. Of particular importance is the confirmation that metastable states with extensive configurational entropy $\Sigma$ play an important role, as revealed by the behaviour of the Franz-Parisi potential \cite{FP-potential}. The existence of a non-trivial point-to-set length $\ell$ and its growth when $\Sigma(T)$ is decreased are now firmly established by numerical simulations \cite{Cavagna,dzero,franz,ceiling,scalliet}. Microscopic calculations have revealed a deep link between RFOT and the Random Field Ising Model \cite{wolynes-rfim,franz-rfim,RFIM1,RFIM2}, with original predictions about the critical behaviour of the Franz-Parisi potential that are in surprisingly good agreement with numerical simulations \cite{guiselin}. 

In the first part of this short survey, we will review recent investigations that broadly support all static aspects of RFOT. Whether or not these thermodynamic features are related to the abrupt dynamical slowdown of super-cooled liquids is still actively debated.
The ``elastic picture”, for
instance, proposes that the chief physical ingredient driving the glass transition is the growth of the plateau shear modulus, $G_{\text{hf}}$, which makes even local moves progressively more difficult. The growth of the activation barrier to flow would then simply mirror the growth of $G_{\text{hf}}$ \cite{shoving}, without having to invoke any growing glassites. In such a picture, a growing static lengthscale is not a crucial ingredient to understand the dynamics of the system \cite{WC}. 
More generally, purely kinetic theories of the glassy slowing down do have forceful advocates. In these theories, the progressive logjam of super-cooled liquids is due to a rarefaction of local ``defects'' that act as facilitators for structural rearrangements \cite{Ritort_Sollich,KCM,Keys}. In this scenario thermodynamics only plays a minor role, or even no role at all. The glass is but a liquid that cannot flow because of kinetic constraints, but there is no driving force towards any kind of locally preferred structure or amorphous order.

In the second part of this paper, we will discuss 
some recent theoretical and numerical arguments raised against the simplest dynamical version of RFOT. It has indeed become indisputable that facilitation (i.e. a local transmission of the activity from one region of space to another) is at play in deeply super-cooled liquids and that such an ingredient, while not in contradiction with RFOT, needs to be incorporated in a more elaborate version of the theory. Conversely, it is difficult to imagine that the universal Adam-Gibbs correlations between static quantities (configurational entropy, jump of specific heat) and dynamical quantities (super-Arrhenius relaxation time, fragility) are only incidental. Furthermore, experimental results on the non-linear dielectric response of glasses cannot be accounted for within thermodynamic-free Kinetically Constrained Models (KCM) \cite{BBL}. In view of the success of RFOT theory in describing the non-trivial nature of the free-energy landscape of glasses, 
and in providing a general theoretical framework able to explain the broad phenomenology of glass-forming liquids,
we believe that the ``final'' theory (if such a thing makes any sense!) will be an appropriate synthesis between RFOT ideas and schematic models of dynamical facilitation. 

Needless to say, the present paper has no ambition to be comprehensive, but rather to give a concise and personal account of the recent successes and difficulties of RFOT theory, together with some open issues. For more exhaustive reviews, in particular regarding the description of the phenomenology of glass-forming liquids by RFOT theory, see \cite{wolynes-review,RFOT2,bbreview}

\section{Statics: Recent Successes}
\label{sec:statics} 

The order parameter of the Random First Order Transition is the overlap (or similarity) $q$ between two typical configurations of the system. By comparing the spatial arrangements in different configurations one can identify whether or not the system is developing amorphous order, even when such arrangements look devoid of any specific ordering and random, like in the liquid state.

More precisely, in the description of the glass transition underpinning RFOT theory one draws an equilibrium configuration $\mathcal{C}_0$ from the Boltzmann measure at temperature $T$ and computes the probability that a second configuration $\mathcal{C}$ drawn the same measure has an overlap $q$ with $\mathcal{C}_0$. The precise definition of the overlap is not important, and several have been used in the literature; what matters is that $q$ measures how much the density fields of the two configurations resemble each other. When $q=0$, the two configurations are completely uncorrelated, like two typical configurations in the liquid state. 

The construction of a consistent statistical theory of the overlap fluctuations in glassy systems has been progressing steadily since the first attempts in the late 80's with a recent acceleration after the exact solution of hard spheres in the limit of infinite dimension was derived \cite{zamponi_book}. It is fair to say that the static description of the RFOT order parameter $q$ and of its spatial fluctuations is by now well understood, as we now briefly review. 

\begin{figure}[!h]
\begin{center}
\includegraphics[width=0.9\linewidth]{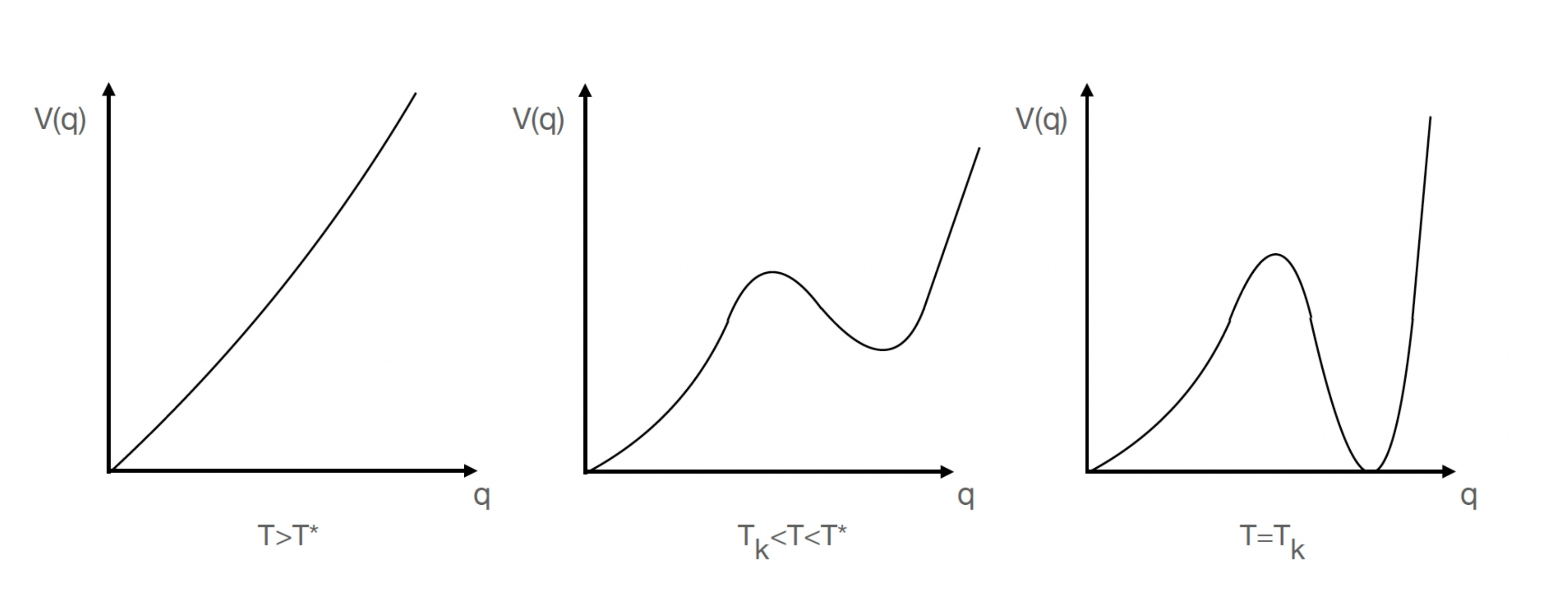}
\caption{Sketch of the mean-field shape of the Franz-Parisi potential in three different temperature regimes.}
\label{fig:potential}
\end{center}
\end{figure}

The free-energy cost associated to an overlap $q$ homogeneous in space is called Franz-Parisi potential $V(q)$ \cite{FP-potential}, and plays the same role of the Landau free-energy in standard critical phenomena. It has different shapes depending on $T$, as shown in Fig. \ref{fig:potential}. Within mean-field theory, for $T>T^\star$, $V(q)$ has only one minimum at $q=0$, corresponding to a liquid; for $T_K<T<T^\star$ a secondary minimum at $q^\star$ appears; finally at $T=T_K$ the secondary minimum $V(q^\star)$ reaches the same value as that of the liquid $V(q=0)$. The existence of the secondary minimum signals the emergence of metastable states, with a non-negligible probability that 
configuration $\mathcal{C}$ resides in the same metastable state as configuration $\mathcal{C}_0$. In this framework, the free-energy difference $V(q^\star)-V(q=0)$ is the entropic cost to enforce that $\mathcal{C}$ is in the same metastable state as $\mathcal{C}_0$. This is the definition of the \textit{configurational entropy} (or complexity) $\Sigma(T)$ -- since one chooses one metastable state over many other possibilities. At $T_K$, this free-energy cost vanishes, and the average of $q$ becomes different from zero: the degeneracy of amorphous ground state becomes sub-extensive. 

At the mean-field level, the statistical theory of overlap fluctuations is similar to the one of a first-order transition. But physical arguments and analytical computations indicate that in finite dimensions the effective theory for the overlap fluctuations is more complex. One of the main insight of Wolynes et al. \cite{KTW3} was indeed that in finite dimensions an extensive configurational entropy necessarily implies the breakup of metastable states into a mosaic of glassites of finite extension $\ell$. As proposed in \cite{Bou04,RFOT2}, the argument leading to $\ell$ can be best understood by considering a \textit{gedanken} cavity of radius $R$ where particles are free to move, when particles outside the cavity are frozen in a typical configuration of the super-cooled liquid. The configuration inside the cavity can be in one among $\exp(R^d \Sigma)$ metastable states. One metastable state (or a few of them) is ``well matched'' with the frozen boundaries and gains a surface energy $\Upsilon R^\theta$ with respect to all the others, with $\theta \leq d-1$. When such a pinning force is not enough to counterbalance configurational entropy, it makes no sense (thermodynamically) to think the inside of the cavity as frozen in a single metastable state. This occurs when $R \gtrsim \ell$, with 
\begin{equation} \label{eq:ell_pts}
    \ell = \left(\frac{\Upsilon(T)}{T \Sigma(T)}\right)^{\frac{1}{d - \theta}}.
\end{equation}
Note that $\ell$ diverges when $\Sigma(T) \to 0$. Indeed, when there is a sub-extensive number of metastable states, it is not contradictory to assume that the whole system is trapped in one of them. The length $\ell$ resembles the nucleation length in first-order phase transition, however one should not forget that the situation is not the standard one since the system is ``nucleating" a metastable state that is statistically identical to the first one. Here, the driving force is the configurational entropy, as there are so many choices of new metastable states to fit in. 

Another important additional physical ingredient for the theory of overlap fluctuations in finite dimensions is that one has to take into account the presence of quenched disorder. 
The spatial randomness is self-generated and arises from the spatial heterogeneity of the metastable states \cite{wolynes-rfim,franz-rfim,RFIM1,RFIM2}. Therefore, the theory 
becomes that of a first-order transition in presence of quenched disorder -- akin to the one of the Random Field Ising Model (RFIM) in an external field -- with a free-energy cost associated to spatial gradients of the overlap $q^\star$. This mapping to the RFIM \cite{wolynes-rfim,franz-rfim,RFIM1,RFIM2} however only holds for static fluctuations. The dynamics associated to overlap fluctuations is much more involved and not fully understood yet. 
\begin{figure}[!h]
\begin{center}
\includegraphics[width=0.3\linewidth]{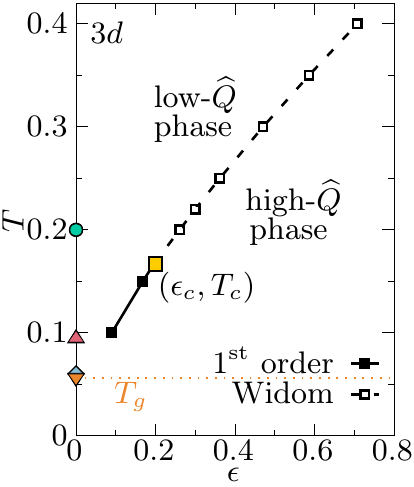}
\includegraphics[width=0.6\linewidth]{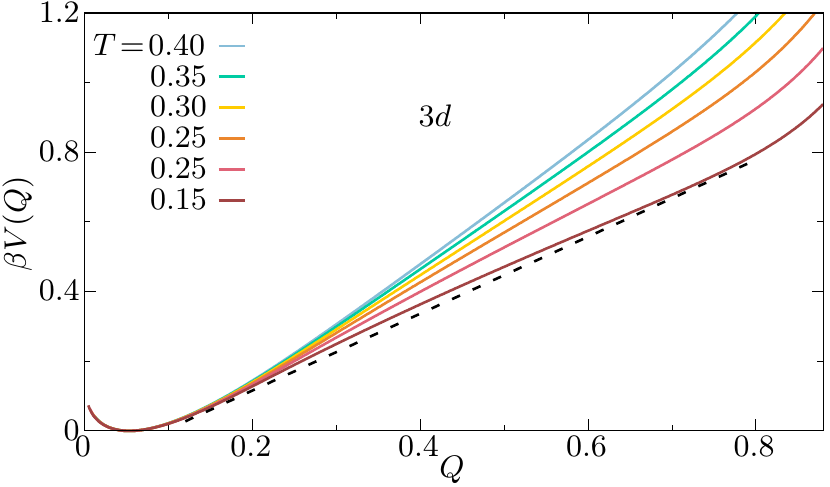}\caption{Left panel: Phase diagram of the glass-forming liquid studied in Ref. \cite{guiselin} in the $(\epsilon,T)$ plane. A critical point (full yellow square) terminates the line of first-order transition (full line) and above it a Widom line is displayed as a dashed line. The authors of Ref. \cite{guiselin} define several characteristic temperatures: the onset temperature of glassy behaviour (green disk), the mode-coupling crossover temperature (pink up triangle), the extrapolated calorimetric glass transition temperature $T_g$ (orange down triangle). Right panel: 
Franz-Parisi (FP) potential of the glass-forming liquid studied in Ref. \cite{guiselin} rescaled by the temperature at several temperatures $T$. The potential is convex at high temperatures, whereas at lower temperature it tends (for large system sizes) to the Maxwell construction of the middle panel of Fig. \ref{fig:potential}. See \cite{guiselin} for more details.}
\label{fig:epsilonphasediagram}
\end{center}
\end{figure}

Elaborating on such a theoretical framework leads to several specific predictions. The first one is that by adding an external ``field'' $\epsilon$ that favours configurations $\mathcal{C}$ with a large overlap with $\mathcal{C}_0$, i.e. adding a linear term $-\epsilon q$ to the free energy, one can tilt the curves in Fig. \ref{fig:potential} and induce a genuine transition towards a non-zero $q$ phase even when $T>T_K$. The theoretical analysis also indicates that by applying the field $\epsilon$ the induced transition changes nature from its $\epsilon=0, T=T_K$ counterpart, and becomes a \textit{bona fide} first-order transition also from a dynamical point of view. 

These and other results offer a highly non-trivial set of predictions, in which the phase diagram acquire an additional $\epsilon$-axis, with new critical behaviours. 
Thanks to the advance in equilibration techniques of atomistic model of glass-forming liquids, these predictions can nowadays be  quantitatively tested. Several such investigations have been performed in the last decade. By and large, numerical results have fully confirmed the theoretical predictions, including the mapping to the RFIM \cite{cardenas,camma-pot,seoane,guiselin}. As an example, we reproduce in Fig. \ref{fig:epsilonphasediagram} the $\epsilon-T$ phase diagram and the corresponding Franz-Parisi potential obtained in \cite{guiselin}, which has indeed the shape expected from mean-field theory in the high and in the intermediate temperature regimes. In the latter regime, one finds that the shape of $V(q)$ precisely corresponds to the Maxwell construction applied to the mean-field result, as expected due to finite dimensional fluctuations. In this temperature regime, the left panel shows the predicted first order transition line in the $\epsilon-T$ plane between a high $q$ and a low $q$ phase. All in all this is quite a remarkable confirmation of the main thermodynamical tenets of RFOT theory. 

Another rather non-trivial test of the theory can be carried out, by pinning a random fraction $c$ of particles in an otherwise equilibrated configuration. The theoretical results of \cite{pinning-camma}, based on RFOT theory, predict 
that the liquid formed by the remaining free particles has a lower configurational entropy than its $c=0$ counterpart. As a consequence one can induce an equilibrium glass transition by increasing the fraction $c$ of pinned particles instead of lowering temperature $T$; the configurational entropy is predicted to vanish at a certain pinning fraction $c_K(T)$, which increases with $T$ ($c_K(T_K)=0$). Moreover, it can be shown that the configuration formed by the unpinned particles remains an equilibrium configuration of the unpinned system, thus allowing one to obtain equilibrated configurations even very close to the glass transition temperature. This opens the way to the study of equilibrium properties very close to $c_K(T)$. 

\begin{figure}[!h]
\begin{center}
\includegraphics[width=0.45\linewidth]{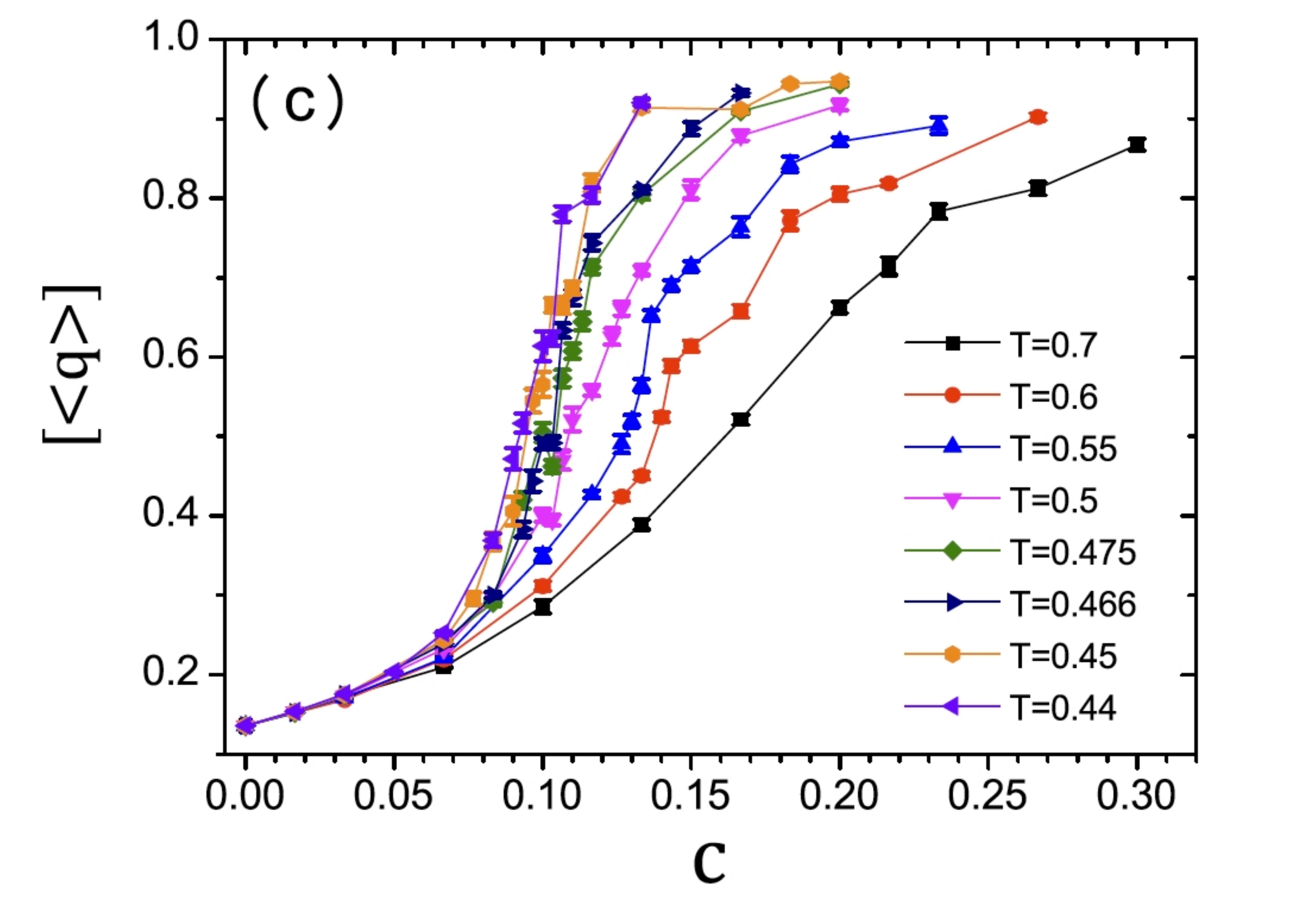}
\includegraphics[width=0.45\linewidth]{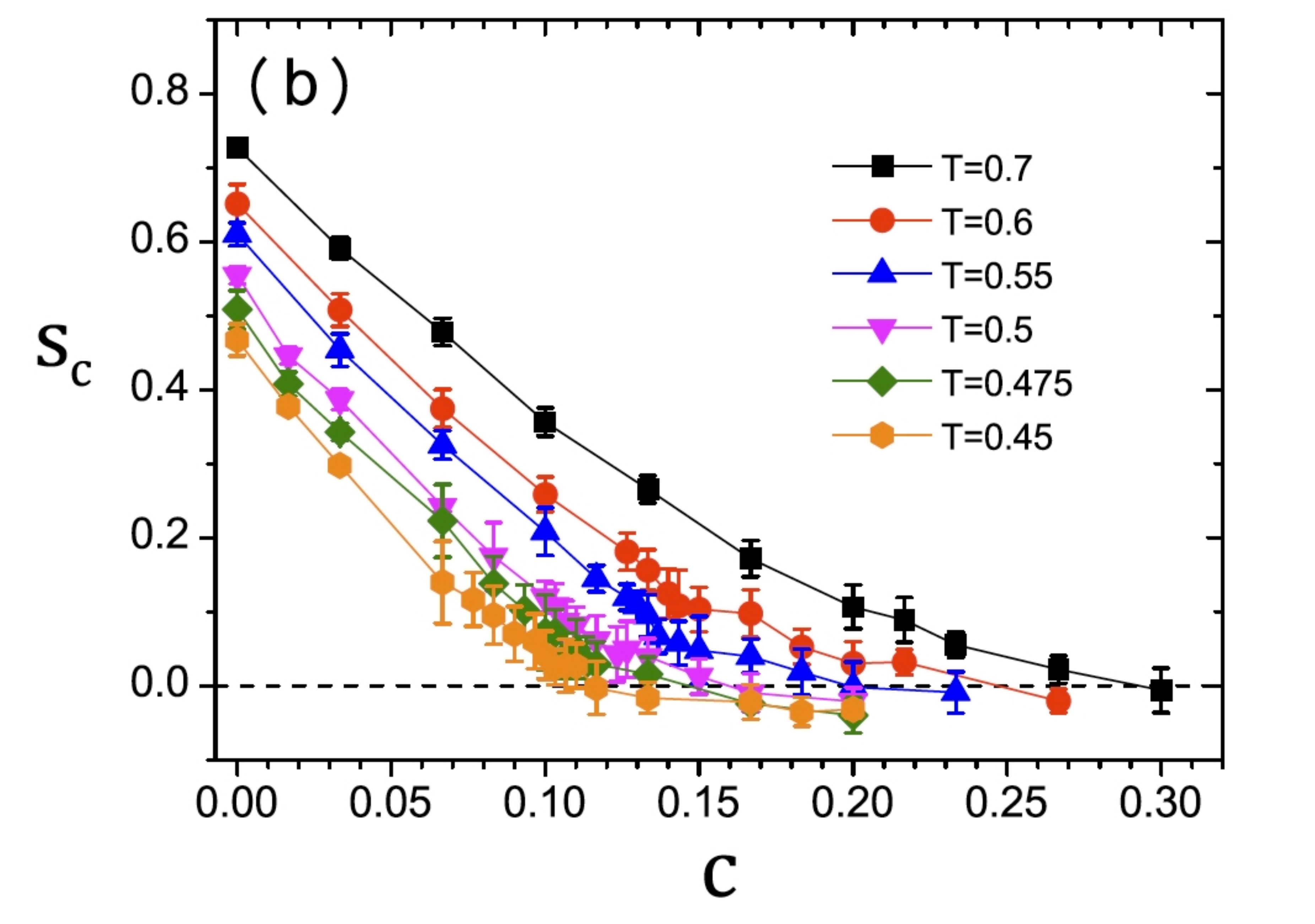}\caption{Left panel: Average overlap as a function of $c$ for different temperatures in the glass-forming liquid studied in \cite{kob}. 
Right panel: Configurational entropy as a function of $c$ for different temperatures for the same system. See \cite{kob} for more details.}
\label{fig:pinning}
\end{center}
\end{figure}

Several numerical and some experimental studies have investigated the behaviour of pinned glass-forming liquids \cite{berthier-pinning,kob,pinning-expt}. Again, theoretical predictions are broadly confirmed, bolstering even further the thermodynamical underpinnings of RFOT theory for finite dimensional glass-forming liquids.
As an example, we show two results from the thorough studies performed in Ref. \cite{kob}. In the left panel of Fig. \ref{fig:pinning} we show 
the average overlap $q$ as a function of the pinning fraction $c$ for different temperatures. Below a certain temperature the increase of $q$ with $c$ is compatible with the existence of an ideal glass phase transition (see  \cite{kob} for more details). In the right panel, the configurational entropy is shown as a function of $c$ for the same range of temperatures, demonstrating that indeed $\Sigma$ vanishes at $c_K(T)$, where $c_K(T)$ increases with temperatures. 

Finally, let us mention the long-standing numerical and experimental efforts to characterize the point-to-set length $\ell$ which measures the spatial extent of amorphous order \cite{Bou04,Montanari-Semerjian}. The first direct numerical evidence of its growth was obtained in 2008 in Refs. \cite{cavagna0,Cavagna}. Thanks to the recent advance in numerical simulations \cite{swap}, the point-to-set length growth and its relation to the decrease of the configurational entropy $\Sigma$, as suggested by Eq. \eqref{eq:ell_pts}, were fully confirmed and quantified in two and three dimensional glass-forming liquids \cite{pts1,pts2}, with an estimate of exponent $\theta$ around $1$ for $d=2$ and $1.5$ for $d=3$, not far from the early predictions of \cite{KTW3}. Interestingly, in the $d=2$ case it was found that $T_K\simeq 0$, i.e. the divergence of the point-to-set length does not seem to take place at finite temperature in agreement with theoretical results suggesting that no glass transition should take place in $d < 3$ \cite{RFIM1,RFIM2}. Attempts to measure static heterogeneities in point-to-set lengths and configurational entropies, and using those to infer the properties of the ``surface tension'' $\Upsilon(T)$ in Eq. \eqref{eq:ell_pts}, are reported in \cite{guiselin_overlaps}.

As far as experiments are concerned, direct measurement of $\ell$ in molecular glass formers is obviously very difficult, see e.g. \cite{pinning-expt,Das0,Das}. An indirect method, based on the idea that frozen glassites respond \textit{collectively} to an oscillating field, is based on non-linear dielectric susceptibilities \cite{Bou05}. If such glassites are compact, theory predicts that the k$^{\text {th}}$-order dielectric susceptibility $\chi_k$ should peak at a value proportional to $\ell^{3(k-1)/2}$, \footnote{Note that the linear dielectric susceptibility ($k=1$) is therefore expected not to show any anomalous increase, as in spin-glasses and in agreement with experiments.} i.e. $\ell^3$ for the third-order non-linear susceptibility and $\ell^6$ for the fifth-order susceptibility, both of which having between measured by two experimental groups \cite{Alb16} (see also O. Dauchot, F. Ladieu, P. Royall  this volume, and references therein). The peak is expected to be located at a frequency $\omega \sim \tau^{-1}$, since at higher frequencies only small clusters can follow the field and at lower frequencies glassites have relaxed and the collective response of frozen dipoles is lost. These predictions agree quantitatively with experiments \cite{Alb16}, which can be seen as the best indirect experimental evidence to date of the growth of a static length scale in super-cooled liquids close to the glass transition. Note that purely kinetic theories of the glass transition, where thermodynamics is trivial or plays no role, cannot explain such anomalous non-linear effects -- see \cite{BBL} for a more detailed discussion, and \cite{speck} for a dissenting view. Of course, the possibility that the RFOT picture explains well all static effects but completely misses the key ingredient that slows down the dynamics is still an open possibility, as we expand on in the next section.  

\section{Dynamics: Recent Difficulties}
\label{sec:dynamics}
As reviewed in the previous section, many static predictions of RFOT are confirmed, sometimes in a non-trivial way, by recent numerical simulations. In particular, the appearance of locally metastable states (``glassites'') of size $\ell$ limited by configurational entropy is now well established. However, the role played by such a static length scale in the dramatic slowing down of super-cooled liquids have recently been the subject of renewed qualms. 

Let us first recall the argument relating $\ell$ to the relaxation time of the liquid. Consider the situation of particles confined in a cavity with frozen amorphous boundary conditions. When the cavity
radius $R$ is less than $\ell$, the liquid inside the cavity is frozen
too, in the sense that only a small subset of configurations
has a significant weight in the Boltzmann measure. When
$R > \ell$, on the other hand, the number of metastable configurations
becomes so large that even when most of them
are incongruous with the boundary conditions, the cavity is driven by entropy into the liquid state. In other words, relaxation of the density field cannot occur unless the radius of the cavity is of the order of, or larger than $\ell$. Note that this statement is independent of the actual dynamics driving the system (provided of course it obeys detailed balance). 

Within the RFOT scenario, the free-energy barrier $B$ for rearrangements in such a cavity of size $\ell$ is argued to grow as
\begin{equation} \label{eq:Barrier}
   B \sim \Delta(T)\, \ell^\psi 
\end{equation}
where $\Delta(T)$ is a temperature dependent energy scale and $\psi$ is a certain exponent.
Associating the relaxation time $\tau$ of the liquid with such a
minimal barrier for decorrelation, one finds, using Eq. \eqref{eq:ell_pts}
\begin{equation} \label{eq:Tau}
    \log \tau \sim \frac{\Delta(T)}{T} \left(\frac{\Upsilon(T)}{T \Sigma(T)}\right)^\alpha, \qquad \alpha = \frac{\psi}{d - \theta}.
\end{equation}
Such an argument is the main claim to fame of RFOT, since it naturally accounts for both (a) the empirically observed Adam-Gibbs correlation between configurational entropy and relaxation time and (b) the strongly non-Arrhenius, Vogel-Fulcher-type increase of $\tau$ in fragile liquids. Experimental and numerical data suggest that in $d=3$, the exponent $\theta$ is around $3/2$, as mentioned in the previous section, whereas the exponent $\psi$ is surprisingly small, $\psi \lesssim 1$ \cite{Cam09}. Ozawa et al., in particular, have revisited experimental data and suggest $\alpha \approx 0.5 \pm 0.2$ for a broad range a materials and model glasses \cite{ozawa}. Even if not a smoking gun proof, a compelling numerical study of L. Berthier \cite{berthiernew} suggests that the Adam-Gibbs correlation holds even locally, therefore establishing a strong link between static and dynamical properties of glasses. 

However, the whole RFOT picture has been challenged by the efficiency of SWAP algorithms to speed up the dynamics \cite{swap,WC}. If local swaps of particles of different radii can decrease the relaxation time by orders of magnitude, doesn't this suggest that kinetic constraints, rather than thermodynamical barriers, dominate the dynamics of glasses at low temperatures? \footnote{An important assumption here is that the atomistic liquids for which the swap algorithm works are representative of glass-forming liquids. In fact, it is certainly possible to construct non-realistic models for which a swap-like algorithms work but that do not have the phenomenology of glass-forming liquids, and to which one should therefore not apply RFOT theory. There are no indication so far that this is the case for the atomistic models studied by swap.} A related argument concerns the Stokes-Einstein decoupling between self-diffusion and collective relaxation, which should be much stronger than experimentally observed if local kinetic constraints were not the dominant effect.\footnote{The self-diffusion constant is $\sim 10^3$ larger than inferred from the value of $\tau$ at $T=T_g$. If particles where individually free to move but collectively trapped, this enhancement factor should be closer to $10^{15}$ \cite{WC,crumbling}.}

Furthermore, facilitation effects have been evidenced and quantified in numerical simulations \cite{candelier,Keys,chacko}, in particular see Ref. \cite{scalliet_guiselin} which explores low temperatures and long-time scales. These numerical studies have clearly established that activity in a certain region of space induces (or ``facilitates'') activity in neighbouring regions. Since this effect is the key ingredient of Kinetically Constrained Model, but is essentially absent from the RFOT description, one is entitled to question the thermodynamic origin of the glass transition. 

Although these arguments cannot be brushed aside lightly, we tend to believe that RFOT can indeed be extended to accommodate all these apparent contradictions. First, let us note that RFOT relies on the existence and proliferation of \textit{local metastable states} below some temperature $T^\star$, which is often identified with the Mode-Coupling temperature. In mean-field, such ``metastable'' states can be defined unambiguously, independently of the dynamics, because energy barriers separating them are infinite. But the chief theoretical difficulty posed by the glass transition precisely lies in correctly handling the concept of metastability in non-mean-field situations. The fact that a collection of micro-states forms a metastable state now depends both on the dynamical rules and on a timescale. This timescale should be long enough to allow 
equilibration among the given set of micro-states and yet be short enough for not allowing the system to escape from that set. Importantly, such a separation of timescales may hold for one set of dynamical rules and not for another. 

\begin{figure}[!h]
\centering
\includegraphics[width = 10cm]{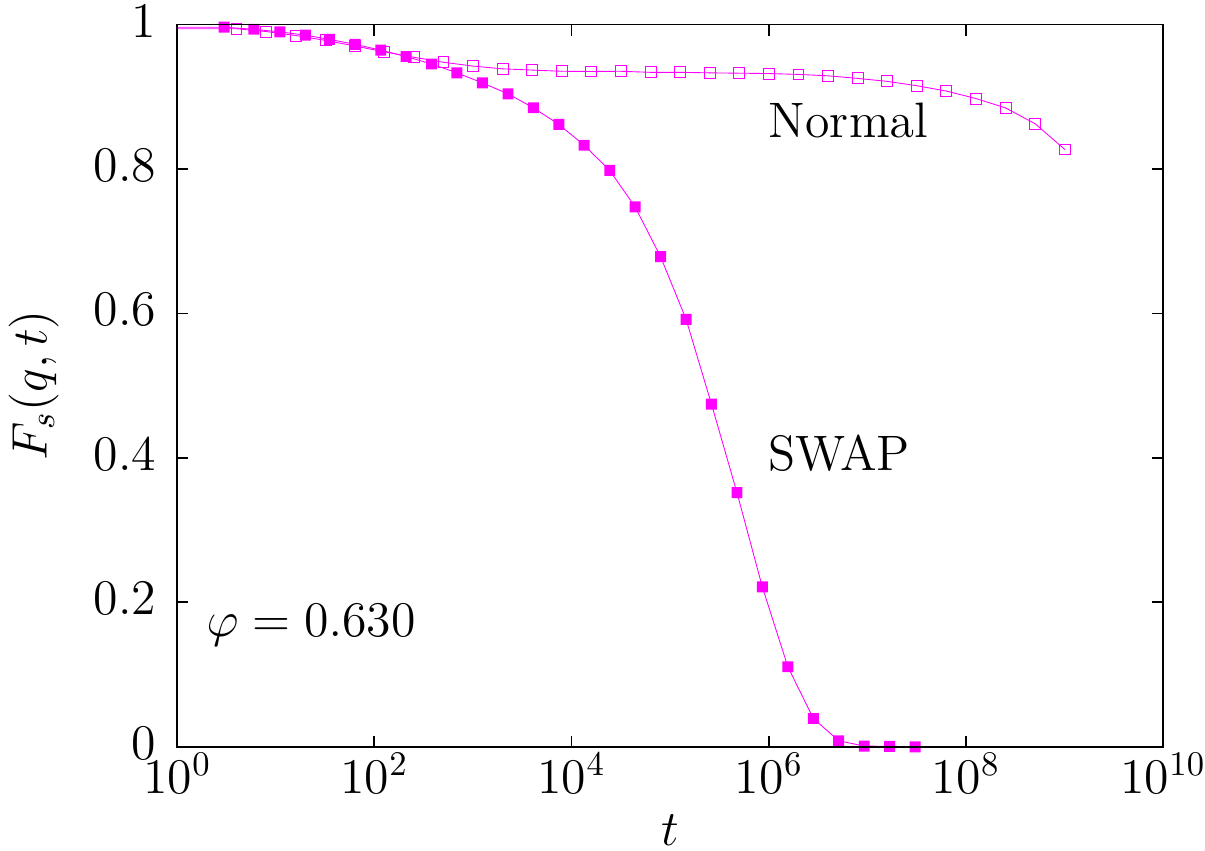}
\caption{Self-intermediate scattering function for a three dimensional polydisperse hard-sphere system, with or without SWAP dynamics. This clearly illustrates the idea of ``crumbling'' metastability: the plateau corresponding to local rigidity -- that extends over more than $4$ decades in time without swaps -- completely disappears in the presence of swaps. From \cite{crumbling}.}
\label{figcorr}
\end{figure}

Now, one can only speak about activation barriers and slow dynamics if the system is locally stable, i.e. if some local rigidity sets in and prevents free flow (on this topic see e.g. \cite{coslovich}). A minimum requirement for local metastability is that the Hessian of the configuration energy computed inside a cavity of size $\ell$ should be definite positive. Now, the SWAP algorithm effectively allows the radius of the particles to fluctuate, thereby increasing the number of degrees of freedom and the dimension of the Hessian matrix. Some unstable directions can therefore appear, that would not exist without swaps. Hence, states that are metastable without swaps can lose their local rigidity when swaps are allowed. A signature of this ``crumbling metastability'' \cite{crumbling} can be seen in Fig. \ref{figcorr}: the two-step relaxation curve, signalling the formation of local cages, is completely wiped out by swaps.  

The excess of unstable directions means that the appearance of local metastability is pushed to lower temperatures, i.e. $T^\star_{\text{swap}} < T^\star$. Even when the point-to-set length $\ell$ is independent of the dynamics and therefore still exists between $T^\star_{\text{swap}}$ and $T^\star$, there is no collective activation barrier in this regime, i.e. $\Delta(T)=0$ (see Eq. \eqref{eq:Barrier}). Below $T^\star_{\text{swap}}$ and for large enough $\ell$, one expects that the (free-) energy barriers given by Eq. \eqref{eq:Barrier} are independent of the dynamics. Hence the relaxation time $\tau_{\text{swap}}(T)$ should in fact massively increase below $T^\star_{\text{swap}}$ to catch up the no-swap value $\tau(T)$. A consequence of this picture is that swap dynamics should lead to anomalously fragile behaviour at low temperature.\footnote{Conversely, adding extra constraints on the dynamics, as proposed by Brito, Kurchan \& Wyart, should reduce the fragility of the glass former.} Hints of such an increased fragility can be seen in Ref. \cite{crumbling}, Fig. 2, bottom graph. The above scenario, that explains the success of the SWAP algorithm in terms of a downward shift of the Mode-Coupling temperature, has been proposed in different incarnations in Refs. \cite{ikeda,brito,szamel,crumbling}. 

Let us now turn to dynamical facilitation. The basic ingredient of all Kinetically Constrained models is that local motion is only possible if a point defect (``facilitator'') passes by. The concentration of these point defects is assumed to be given by $\rho \sim e^{-J/T}$ where $J$ is a large activation energy. Furthermore, these point defects slowly propagate in space, with a distance travelled growing like $r \sim t^{1/z}$ where $z$ is a temperature dependent dynamical exponent. When $z > d$, exploration of space is compact. Therefore, all regions have had a chance to decorrelate when $\rho r^d \sim 1$, which translates into a relaxation time $\tau_{\text{kcm}}$ given by \cite{Ritort_Sollich, Keys}
\begin{equation}
    \tau_{\text{kcm}} \sim e^{\frac{Jz(T)}{dT}}.
\end{equation}
In some models, like the ``East'' model, the temperature dependence of the dynamical exponent is given by $z(T)=T_0/T$, which in turn leads to a super-Arrhenius growth of the relaxation time \cite{Keys}. In this picture, there is no thermodynamical correlations between particles (i.e. no ``amorphous order''), nor any static inhomogeneities. Glassy slowing down is the result of two effects: the rarefaction of facilitators, and their more and more sluggish progression in space. Such a picture accounts for the growth of dynamical heterogeneities when the temperature is reduced: the size of the regions that relax in a correlated way is given by the typical equilibrium distance between defects, i.e. $\xi_{\text{dyn}} \sim e^{J/dT}$. 

Although facilitation is \textit{not} a key ingredient explaining glassy slowdown in the context of RFOT, it has always been assumed that some kind of facilitation is implied by the ``mosaic'' picture of glassites (see, e.g. \cite{Xia,WW,crumbling}). Indeed, the thermodynamic argument balancing configurational entropy and boundary mismatch energy only makes sense, strictly speaking, when the particles outside the cavity are frozen. However, in a bulk liquid, boundary conditions acting on this \textit{gedanken} cavity are evolving on the very same timescales as the cavity itself. It is clear that the rearrangements of the nearby glassites, by affecting those boundary conditions, can increase or decrease the energy barrier of the central glassite. 
Hence, through these boundary conditions, some activity is transferred from one region of size $\ell$ to the neighbouring regions. A concrete mechanism for this kind of facilitation, based on elastic interactions, has been put forward and observed in numerical simulations \cite{chacko}.

The naive RFOT argument leading to Eq. \eqref{eq:Tau} should therefore be made self-consistent to account for facilitation. This bootstrap effect is presumably responsible for (a) some dynamical correlations extending over regions much larger than $\ell$; (b) some acceleration of the dynamics compared to the case of frozen boundary
conditions. In fact, as we have argued in \cite{crumbling}, the effective barrier energy $\Delta(T)$ could even renormalise to zero in a certain range of temperatures for large enough coupling between different regions.

Be that as it may, the relaxation timeline suggested by RFOT + facilitation is the following: glassites with exceptionally low local energy barriers (but still scaling as Eq. \eqref{eq:Barrier}) start being activated. The activity is progressively transferred to neighbouring regions through local facilitation. Once these coarsening domains have met and invaded the whole system, full relaxation is achieved. Such a scenario, also suggested in \cite{scalliet_traps}, seems quite compatible with the recent numerical results of \cite{scalliet_guiselin}. Although superficially similar, we note several important differences with the KCM scenario: (a) the analogue point defects (which we would like to call ``activons'') are not conserved but can spontaneously appear, branch off or disappear; (b) these activons propagate in a strongly heterogeneous environment: glassites with particularly high barriers may act as obstacles hampering their progression; (c) the super-Arrhenius growth of the relaxation time still primarily comes from the growth of the local energy barriers Eq. \eqref{eq:Barrier}, and \textit{not} from the anomalously slow propagation of point defects, as is the case in e.g. the East model.  

Of course the impressionist picture outlined above needs to be made more quantitative, and time will tell whether one can indeed reconcile the dynamical predictions of RFOT with all recent numerical results that have allowed one to get an unprecedented level of information about how the dynamics of deeply super-cooled liquids unfolds ``in the first 30 milliseconds'' \cite{scalliet_guiselin}. In particular, the relative importance of local activation vs. sluggish facilitation is more than ever a key issue.

\section{Discussion \& Open Issues}

As reviewed in section \ref{sec:statics}, many rather non-trivial predictions of RFOT theory about the \textit{statics} of super-cooled liquids have been confirmed by analytical calculations  or numerical simulations in the last decade (see e.g. \cite{zamponi_book, guiselin}). Still, some aspects would be well worth further investigations. An experimental confirmation that super-cooled liquids can become glasses by confining them in small enough frozen cavities formed by the same liquid and/or when pinning a high enough fraction of particles would be a vindication of the basic premises of RFOT. Although challenging (see e.g. \cite{pinning-expt,Das0,Das}), such experiments would provide a direct measure of the point-to-set length $\ell$, for which we only have indirect evidence, for example from the behaviour of non-linear dielectric susceptibilities. 

On another front, there are still very detailed RFOT predictions concerning the RFIM effective theory for static overlap fluctuations that remain to be tested. One is about the fluctuations of the interfaces separating glassites. It was theoretically shown in \cite{surface-camma} that these fluctuations coincide with the one of domain walls in the Random Field Ising Model, as anticipated in \cite{KTW3}. This leads to the appearance of an additional length-scale $\ell_\perp \sim \Sigma(T)^{-1/2}$ associated to the wandering of such interfaces. Such a prediction could be tested by focusing on a pinning set-up in which all particles on one side of a plane are pinned, and the fluctuations of the overlap field on the other side are studied numerically. Another important step is measuring the key observables that would allow one to identify quantitatively the effective RFIM-like theory of the overlap fluctuations. First measurements of the ``surface tension'' $\Upsilon(T)$ were obtained in \cite{guiselin_overlaps}. But another important piece of information that is currently missing is about the nature of the Franz-Parisi potential fluctuations. The spatial covariance of these fluctuations would allow one to obtain the strength of the random field term in the effective RFIM description. This is a crucial aspect, since too large a random field could destroy the transition (as it is known for the RFIM), thus predicting the absence of any RFOT in some glass-forming liquids \cite{wolynes-rfim,RFIM1,RFIM2}, which would then have a very different phenomenology.

Concerning dynamics, the situation is more subtle. Whereas the RFOT scenario accounts for most of the phenomenology of the glass transition and is ruled out by no experiment nor numerical simulation, there are a number of loose ends that we feel need to be tied up before victory can be declared. A major aspect is facilitation, as emphasized in section \ref{sec:dynamics}. Is the growth of the relaxation time fully explained by the growth of the point-to-set length and the corresponding energy barriers of elementary glassites, or is the slowdown the result of a subtle interplay between local events and spatial propagation of activity? Why is the growth of relaxation time so closely related to the increase of the high frequency shear modulus $G_{\text{hf}}$, as indeed predicted by elastic theories \cite{shoving, WC}? If we believe in glassite activation, why is the barrier exponent $\psi$ extracted from numerical simulations \cite{Cam09} and experiments \cite{ozawa} systematically on the low side ($\psi \lesssim 1$) when one would naively expect $\psi \geq \theta$, with $\theta \gtrsim 3/2$ in three dimensions? 

From an experimental standpoint, we wish to suggest two possibly interesting directions:
\begin{itemize}
    \item 
One is to try to interpret the high-frequency power-law behaviour of the non-linear susceptibilities $\chi_3(\omega)$,  $\chi_5(\omega)$ measured in \cite{Alb16}, in the regime corresponding to the  ``excess wing'' for linear susceptibilities (see also \cite{Bau13}). If we follow the interpretation of \cite{scalliet_guiselin}, the excess wing is due to the early activation sites that then grow and propagate through facilitation. The corresponding behaviour of $\chi_3(\omega)$,  $\chi_5(\omega)$ could provide some information about the spatial structure of those ``activons''. Experimental measurements of (third harmonic) non-linear mechanical response would also be highly interesting \cite{fuchs, fuchs2}.   
    \item The second concerns rheology and fracture. As we pointed out in ref. \cite{RFOT2}, RFOT theory suggests a strong crossover from a high viscosity regime at low shear stress $\sigma$ to a low viscosity regime at higher shear stress, when the elastic energy $G_{\text{hf}} \sigma^2 \ell^3$ stored in a glassite exceeds the energy barrier $B$ given by Eq. \eqref{eq:Barrier}. Interestingly, the cross-over stress should decrease as temperature decreases \cite{RFOT2}. Similarly, when a fracture propagates inside a super-cooled liquid close to the glass transition, one may also expect a brittle-ductile transition when concentrated stresses at the tip of the crack are able to ``liquify'' the glass ahead of the fracture front, with possibly hysteretic effects -- see Ref. \cite{babs}. The fracture surface left behind should correspondingly reveal an interesting crossover length between two roughness exponents \cite{physrep}. 
\end{itemize}

In conclusion, we hope that this short review will spur further theoretical discussions and more experimental/numerical investigations that will help shed light on a long-standing conundrum: after all, why don't glasses flow? 

\subsection*{Acknowledgments} We want to deeply thank L. Berthier, E. Bouchaud, C. Cammarota, F. Ladieu, M. Ozawa, D.R. Reichman, C. Scalliet, G. Tarjus, M. Wyart for many discussions on all these issues over the years. GB is partially supported from the Simons Foundation (Grant No. 454935).


\begin{thebibliography}{99.}%
%
%


\bibitem{bbreview} L. Berthier and G. Biroli, {Theoretical perspective on the glass transition and amorphous materials} Reviews of modern physics 83.2 (2011): 587.

\bibitem{yoshino} Yoshino, H. (2012). Replica theory of the rigidity of structural glasses. The Journal of Chemical Physics, 136(21), 214108.

\bibitem{KTW1} Kirkpatrick, T. R., \& Thirumalai, D. p-spin-interaction spin-glass models: Connections with the structural glass problem. Physical Review B, 36(10), 5388 (1987).

\bibitem{KTW2} Kirkpatrick, T. R., \& Thirumalai, D. Comparison between dynamical theories and metastable states in regular and glassy mean-field spin models with underlying first-order-like phase transitions. Physical Review A, 37(11), 4439 (1988).

\bibitem{KTW3} Kirkpatrick, T. R., Thirumalai, D., \& Wolynes, P. G. Scaling concepts for the dynamics of viscous liquids near an ideal glassy state. Physical Review A, 40(2), 1045 (1989).



\bibitem{RFOT1} M. Dzero, J. Schmalian, and P. G. Wolynes, in { Structural glasses and super-cooled Liquids: theory, experiment, and applications}, P. G. Wolynes, V. Lubchenko, Eds. (Wiley,  2012), pp. 193-222 


\bibitem{RFOT2} G. Biroli, and J.-P. Bouchaud, in {Structural glasses and super-cooled liquids: theory, experiment, and applications},  P. G. Wolynes, V. Lubchenko, Eds. (Wiley, 2012), pp. 31-114.


\bibitem{Bou04} J.-P. Bouchaud, and G. Biroli, {On the Adam-Gibbs-Kirkpatrick-Thirumalai-Wolynes scenario for the viscosity increase in glasses}, J. Chem. Phys. {\bf 121}, 7347 (2004).

\bibitem{Montanari-Semerjian}
Montanari, A., \& Semerjian, G. (2006). {Rigorous inequalities between length and time scales in glassy systems}. Journal of statistical physics, 125(1), 23-54.



\bibitem{Cavagna} G. Biroli,  J. P. Bouchaud, A. Cavagna, T. S. Grigera,  \& P. Verrocchio. {Thermodynamic signature of growing amorphous order in glass-forming liquids.} Nature Physics, 4(10), 771-775 (2008).


\bibitem{wolynes-review} P.G. Wolynes and V. Lubchenko, (2012). {Structural glasses and super-cooled liquids: Theory, experiment, and applications}. John Wiley \& Sons.

\bibitem{FP-potential} S. Franz and G. Parisi, {Recipes for metastable states in spin glasses}. Journal de Physique I 5.11 (1995): 1401-1415.


\bibitem{dzero} M. Dzero, J. Schmalian, P.G. Wolynes, P. G. {Activated events in glasses: The structure of entropic droplets}. Physical Review B, {\bf 72}(10), 100201 (2005).

\bibitem{franz}
Franz, S. (2005). First steps of a nucleation theory in disordered systems. Journal of Statistical Mechanics: Theory and Experiment, 2005(04), P04001.

\bibitem{ceiling} 
L. Berthier et al. {Configurational entropy measurements in extremely super-cooled liquids that break the glass ceiling}. Proceedings of the National Academy of Sciences, {\bf 114}(43), (2017) 11356-11361.


\bibitem{scalliet}
L. Berthier, M. Ozawa, C. Scalliet, {Configurational entropy of glass-forming liquids}. The Journal of chemical physics, {\bf 150}(16),  (2019) 160902.


\bibitem{wolynes-rfim}
Stevenson, J. D., Walczak, A. M., Hall, R. W., \& Wolynes, P. G. (2008). Constructing explicit magnetic analogies for the dynamics of glass forming liquids. The Journal of chemical physics, 129(19), 194505.


\bibitem{franz-rfim} S. Franz, G. Parisi, F. Ricci-Tersenghi, T. Rizzo, (2011). { Field theory of fluctuations in glasses}. The European Physical Journal E, 34(9), 1-17.


\bibitem{RFIM1}
G. Biroli, C. Cammarota, G. Tarjus, and M. Tarzia, (2018). {Random-field Ising-like effective theory of the glass transition. I. Mean-field models.} Physical Review B, 98(17), 174205.

\bibitem{RFIM2}
G. Biroli, C. Cammarota, G. Tarjus, G. and M. Tarzia, (2018). {Random field Ising-like effective theory of the glass transition. II. Finite-dimensional models.} Physical Review B, 98(17), 174206.


\bibitem{guiselin} B. Guiselin, L. Berthier, \& G. Tarjus, (2022). {Statistical mechanics of coupled super-cooled liquids in finite dimensions}. SciPost Physics, 12(3), 091.

\bibitem{shoving}
J. C. Dyre, T. Christensen, N. B. Olsen. {Elastic models for the non-Arrhenius viscosity of glass-forming liquids}, Journal of non-crystalline solids, {\bf 352}(42-49) (2006). 

\bibitem{WC}
M. Wyart, M. Cates. {Does a growing static length scale control the glass transition?} Physical review letters, {\bf 119}(19), 195501 (2017).


\bibitem{Ritort_Sollich} Ritort, F., \& Sollich, P. Glassy dynamics of kinetically constrained models. Advances in physics, 52(4), 219-342 (2003).


\bibitem{KCM} D. Chandler, and J. P. Garrahan, {Dynamics on the Way to Forming Glass: Bubbles in Space-Time},
Annu. Rev. Phys. Chem. 61, 191-217 (2010).

\bibitem{Keys} Keys, A. S., Hedges, L. O., Garrahan, J. P., Glotzer, S. C., \& Chandler, D. (2011). Excitations are localized and relaxation is hierarchical in glass-forming liquids. Physical Review X, 1(2), 021013.


\bibitem{BBL} Biroli, G., Bouchaud, J. P., \& Ladieu, F. (2021). Amorphous Order and Nonlinear Susceptibilities in Glassy Materials. The Journal of Physical Chemistry B, 125(28), 7578-7586.


\bibitem{zamponi_book} Parisi, G., Urbani, P., \& Zamponi, F. (2020). Theory of simple glasses: exact solutions in infinite dimensions. Cambridge University Press.



\bibitem{cardenas} M. Cardenas, S. Franz, and G. Parisi. {Constrained Boltzmann-Gibbs measures and effective potential for glasses in hypernetted chain approximation and numerical simulations}. The Journal of chemical physics 110.3 (199

\bibitem{camma-pot} C. Cammarota, A. Cavagna, I. Giardina, G. Gradenigo,T.S.  Grigera, G.  Parisi, P. Verrocchio,  (2010). {Phase-separation perspective on dynamic heterogeneities in glass-forming liquids}. Physical review letters, 105(5), 055703.

\bibitem{seoane} G. Parisi, and B. Seoane, (2014). {Liquid-glass transition in equilibrium}. Physical Review E, 89(2), 022309.


\bibitem{pinning-camma} C. Cammarota and G. Biroli, (2012). {Ideal glass transitions by random pinning}. Proceedings of the National Academy of Sciences, 109(23), 8850-8855.



\bibitem{berthier-pinning}
W. Kob and L. Berthier, {Probing a liquid to glass transition in equilibrium}. Physical review letters 110.24 (2013): 245702.

\bibitem{kob} M. Ozawa, W. Kob, A. Ikeda, and K. Miyazaki, (2015). {Equilibrium phase diagram of a randomly pinned glass-former}. Proceedings of the National Academy of Sciences, 112(22), 6914-6919.

\bibitem{pinning-expt} Gokhale, S., Hima Nagamanasa, K., Ganapathy, R., \& Sood, A. K. (2014). Growing dynamical facilitation on approaching the random pinning colloidal glass transition. Nature communications, 5(1), 1-7.

\bibitem{cavagna0} A. Cavagna, T. Grigera and P. Verrocchio (2007). {Mosaic multistate scenario versus one-state description of super-cooled liquids}. Physical review letters, 98(18), 187801.


\bibitem{swap}
A. Ninarello, L. Berthier, D. Coslovich. {Models and algorithms for the next generation of glass transition studies} Physical Review X, {\bf 7}(2), 021039 (2017).

\bibitem{pts1} L. Berthier, P. Charbonneau, D. Coslovich, A. Ninarello, M. Ozawa, S. Yaida, (2017). {Configurational entropy measurements in extremely super-cooled liquids that break the glass ceiling.} Proceedings of the National Academy of Sciences, 114(43), 11356-11361.

\bibitem{pts2} L. Berthier, P. Charbonneau, A. Ninarello, M. Ozawa, and S. Yaida,  (2019). {Zero-temperature glass transition in two dimensions}. Nature communications, 10(1), 1-7.


\bibitem{guiselin_overlaps} Guiselin, B., Tarjus, G., \& Berthier, L. (2022). Static self-induced heterogeneity in glass-forming liquids: Overlap as a microscope. The Journal of Chemical Physics, 156(19), 194503.


\bibitem{Das0} Das, R., Chakrabarty, S., \& Karmakar, S. (2017). Pinning susceptibility: a novel method to study growth of amorphous order in glass-forming liquids. Soft matter, 13(38), 6929-6937.

\bibitem{Das} Das, R., Bhowmik, B. P., Puthirath, A. B., Narayanan, T. N., \& Karmakar, S. (2021). Soft-Pinning: Experimental Validation of Static Correlations in super-cooled Molecular Glass-forming Liquids. arXiv preprint arXiv:2106.06325.


\bibitem{Bou05} J.-P. Bouchaud, and G. Biroli, {Nonlinear susceptibility in glassy systems: A probe for cooperative dynamical length scales}, Phys. Rev. B {\bf 72}, 064204 (2005).


\bibitem{Alb16} S. Albert, Th. Bauer, M. Michl, G. Biroli, J.-P. Bouchaud, A. Loidl, P. Lunkenheimer, R. Tourbot, C. Wiertel-Gasquet, and F. Ladieu, {Fifth-order susceptibility unveils growth of thermodynamic amorphous order in glass-formers}, Science {\bf 352}, 1308 (2016).


\bibitem{speck} 
T. Speck, { Dynamic facilitation theory: a statistical mechanics approach to dynamic arrest}, JSTAT 084015 (2019).


\bibitem{Cam09} C. Cammarota,  A. Cavagna, G. Gradenigo, T. S. Grigera and P. Verrocchio, {Numerical determination of the exponents controlling the relationship between time, length, and temperature in glass-forming liquids}, J. Chem. Phys. {\bf 131}, 194901 (2009).


\bibitem{ozawa} Ozawa, M., Scalliet, C., Ninarello, A., \& Berthier, L. (2019). Does the Adam-Gibbs relation hold in simulated super-cooled liquids?. The Journal of chemical physics, 151(8), 084504.

\bibitem{berthiernew} Berthier, L. (2021). Self-induced heterogeneity in deeply super-cooled liquids. Physical Review Letters, 127(8), 088002.

\bibitem{crumbling} L. Berthier, G. Biroli, J.-P. Bouchaud, G. Tarjus,  { Can the glass transition be explained without a growing static length scale?} The Journal of chemical physics, {\bf 150}(9), 094501 (2019). 

\bibitem{coslovich} Coslovich, D., Ninarello, A., \& Berthier, L. (2019). A localization transition underlies the mode-coupling crossover of glasses. SciPost Physics, 7(6), 077.

\bibitem{candelier} Candelier, R., Widmer-Cooper, A., Kummerfeld, J. K., Dauchot, O., Biroli, G., Harrowell, P., \& Reichman, D. R. (2010). Spatiotemporal hierarchy of relaxation events, dynamical heterogeneities, and structural reorganization in a supercooled liquid. Physical review letters, 105(13), 135702.




\bibitem{chacko}
R. Chacko, F. Landes, G. Biroli, O. Dauchot, A. Liu, and D. R. Reichman (2021). {Elastoplasticity Mediates Dynamical Heterogeneity Below the Mode Coupling Temperature.} Physical Review Letters, 127(4), 048002.


\bibitem{scalliet_guiselin} Scalliet, C., Guiselin, B., \& Berthier, L. (2022). Thirty milliseconds in the life of a super-cooled liquid. arXiv preprint arXiv:2207.00491.


\bibitem{ikeda} Ikeda, H., Zamponi, F., \& Ikeda, A. (2017). Mean field theory of the swap Monte Carlo algorithm. The Journal of chemical physics, 147(23), 234506.

\bibitem{brito} Brito, C., Lerner, E., \& Wyart, M. (2018). Theory for swap acceleration near the glass and jamming transitions for continuously polydisperse particles. Physical Review X, 8(3), 031050.

\bibitem{szamel} Szamel, G. (2018). Theory for the dynamics of glassy mixtures with particle size swaps. Physical Review E, 98(5), 050601.



\bibitem{Xia} X. Xia, \& P. G. Wolynes, P. G., { Microscopic theory of heterogeneity and nonexponential relaxations in super-cooled liquids.} Physical Review Letters, 86(24), 5526 (2001).


\bibitem{WW} Wisitsorasak, A., \& Wolynes, P. G. (2014). Dynamical heterogeneity of the glassy state. The Journal of Physical Chemistry B, 118(28), 7835-7847.


\bibitem{scalliet_traps} Guiselin, B., Scalliet, C., \& Berthier, L. (2022). Microscopic origin of excess wings in relaxation spectra of supercooled liquids. Nature Physics, 18(4), 468-472.

































\bibitem{surface-camma} G. Biroli and C. Cammarota, {Fluctuations and shape of cooperative rearranging regions in glass-forming liquids}. Physical Review X 7.1 (2017): 011011.


\bibitem{Bau13} Th. Bauer, P. Lunkenheimer, S. Kastner, and A. Loidl, {Nonlinear dielectric response at the excess wing of glass-forming liquids}, Phys. Rev. Lett. {\bf 110}, 107603 (2013).



\bibitem{fuchs}
C\'ardenas, H., Frahsa, F., Fritschi, S., Nicolas, A., Papenkort, S., Voigtmann, T., \& Fuchs, M. (2017). Nonlinear mechanical response of super-cooled melts under applied forces. The European Physical Journal Special Topics, 226(14), 3039-3060.

\bibitem{fuchs2}
Seyboldt, R., Merger, D., Coupette, F., Siebenbürger, M., Ballauff, M., Wilhelm, M., \& Fuchs, M. (2016). Divergence of the third harmonic stress response to oscillatory strain approaching the glass transition. Soft Matter, 12(43), 8825-8832.

\bibitem{babs} Gimenes, G. E., \& Bouchaud, E. (2018). Flow and fracture near the sol–gel transition of silica nanoparticle suspensions. Soft Matter, 14(39), 8036-8043.

\bibitem{physrep} Bonamy, D., \& Bouchaud, E. (2011). Failure of heterogeneous materials: A dynamic phase transition?. Physics Reports, 498(1), 1-44.

\end{thebibliography}
\end{document}